\documentclass[twocolumn]{article}
\usepackage{amsfonts,a4wide,graphicx}

\begin{document}

\title{\textbf{RELATIVISTIC CRYSTALLINE SYMMETRY 
BREAKING AND ANYONIC STATES IN MAGNETOELECTRIC
SUPERCONDUCTORS\thanks{MEIPIC - 2, Topic\#: 04\_Rub}}}

\author{Jacques L. RUBIN\thanks{E-mail: jacques.rubin@inln.cnrs.fr}\\
\textit{Institut du Non-Lin\'eaire de Nice (INLN),\/}\\
\textit{UMR 129 CNRS - Universit\'e de Nice - Sophia Antipolis,\/}\\
\textit{1361 route des lucioles, 06560 Valbonne, France\/}}

\date{Enhanced and english corrected unpublished 2002 version of the 1993
one,\\ the latter being published in the\\
\textsc{Proceedings of the international conference on\\  ``Magnetoelectric Interaction
Phenomena in Crystals, Part I"},\\ Ascona  (Switzerland) Sept. 93.\\ \textsc{Ferroelectrics},
Vol. 161(1-4) (1994),  pp. 335-342.\\
\mbox{}\\
\today}

\maketitle

\begin{abstract}
There exists a connection between the
creation of toroidal moments (TM) and the breaking of the one-cell
relativistic crystalline symmetry (RCS) associated to any given
crystal \cite{rubnuovo} into which non-trivial magnetoelectric coupling
effects (ME) exist \cite{schmid73} . Indeed, in this kind of crystals, any
interaction between a charge carrier and an elementary magnetic cell can
breaks the RCS of this previous given cell by varying, in the simplest case,
the continuous defining parameters of the initial RCS.\par This breaking 
can be associated to a change of the initial Galilean proper frame of any
given carrier to an ``effective" one, into which the RCS of the interacting
cell is kept. We can speak of a kind of ``inverse" kineto-magnetoelectric
effect
\cite{ascher73}. The magnetic groups compatible with such process have been
computed
\cite{rubnuovo}. Moreover, one can notice that the TM's break the P and T
symmetries but not the PT one as in anyons theories \cite{wilczek90,dubovik}.
This breaking creates so-called Nambu--Goldstone bosons generating
``effective" magnetic monopoles. These consequences allow us to claim, first,
that anyons are charge carriers associated with ``effective" magnetic
monopoles, both with TM's, and second, that ME can be highly considered in
superconductors theory.
\end{abstract}\bigskip


\section{\normalsize\hskip-.5cm - Quantized Hall effect, high-T$_c$
superconductivity and anyons theory}

In this chapter, we very briefly recall the quantized Hall
effect and the link with both the high-T$_c$ 
superconductivity theory and the anyons theory \cite{wilczek90}.
It is well-known that applying a magnetic field in the orthogonal
direction of a conducting layer crossed over by an electric current,
then an  electric field orthogonal both to the longitudinal
electric current and the magnetic field appears. That is the
so-called classical Hall effect. In addition, a magneto-resistivity
(or magneto-conductivity) effect changes, in particularly, values of
the longitudinal resistivity. 

Then, currents and electric fields are no longer proportionnal but are
related together by a two-dimensional antisymetric resistivity tensor
depending on the applied magnetic field intensity~$B$. Each  diagonal
coefficient  equals the longitudinal resistance, whereas the
non-diagonal coefficients are precisely plus or minus the so-called
Hall resistance $R_H = B/(n e c)$, where $n$ is the
surface density of electric charges. The latter is linearly
depending on the filling factor $\nu = n / n_B$. This
previous relation links the Hall resistance to the number of filled
so-called Landau levels in a bounded conducting layer. These levels
are defined by the stationary eigenfunctions of a one-particle
Schr\"odinger equation in a two-dimensional space with a magnetic
interaction. It is a harmonic oscillator-like equation with
eigenvalues $E_N = \hbar \omega_C ( N + 1/2 )$,
where $\omega_C$ is the cyclotron frequency ($N\in \mathbb{N}$). The boundary
conditions make each Landau level highly degenerate with a 
$n_B$ surface density of degeneracies. 

So, a linear variation of the Hall resistance would have been
experimentally observed, but as K. Von Klitzing and al. \cite{klitzing} have
shown in 1980, the Hall resistance is quantized and the evolution with
the magnetic intensity presents plateaus for integer values of
$\nu$. More, on the plateaus and when the temperature goes to zero,
the longitudinal resistance tends also towards zero. The system becomes
non-dissipative and  permanent longitudinal currents appear.
 
In 1982, D.C. Tsui and al. \cite{tsui} observed the fractional quantized Hall
effect for rational values of $\nu \leq 1$ ($\nu = p/q$ with $q$
odd), meaning that only the first Landau level is excited and that
electron-electron correlations produce condensation in a lower
energy level. Such results have been explained in 1983 by B. B.
Laughlin \cite{laughlin} considering a model of free fermions in a plane,
interacting with a very particular gauge potential (the ``anyonic"
vector potential) not a priori electromagnetic. The
main caracteristic of this potential is the existence of
singularities in the plane at each locations of the fermions.
Solving the corresponding Hamiltonian leads to the so-called
Laughlin function, which points out strong correlations at a finite
distance depending on  the gauge potential intensity. Moreover, the
computation of the total energy of the system effectively shows
singularities (``cusps") for rational values (with an odd
denominator) of $\nu$.  \par
One can display this model in considering excited fractional electric charges
$e/q$, i.e. each cyclotron orbit contains into its
associated  circle, an odd  number $q$ of singularities (or
electrons) of the gauge potential; The latter being viewed as effective
magnetic singular flux tubes. The electric charge is shared by $q$ of
these flux tubes. It is from these considerations that the anyons theory and
the anyonic superconductivity are built.  Indeed, the gauge choice made by
Laughlin, generates a non-trivial Aharonov-B\"ohm phase when each
electron moves on its cyclotron orbit. In particularly, it follows
that the permutation or exchange operator has no longer  the eigenvalues $\pm
1$ only, but more generally, unitary complex numbers. That is the main
property of anyons together with the P and T violations.\par
Actually, the litterature about anyons theory and its mathematical formalisms
and tools, is highly increased. The basic mathematical tool of all of these
formalisms, is to  take into account the so-called Chern-Simon (CS)
Lagrangian term
\cite{wilczek90,eguchi}. The latter breaks, as expected, the P and T
symmetries and is associated to the singularities of the gauge fields. It is
necessary to point out a common confusion about the CS term. It is
absolutely not the CS term in a 2+1 dimension space (it doesn't exist !),
but the one of a curvature field in the 3+1 Minkowski space ! After
integrating  on a 2+1 dimensional hypersurface, bounded by the plane of the
conducting layer, we obtain the ``CS term" of the two-dimensional
anyons theory. Moreover, the 3+1 CS term is ${}^{\ast}\!F\,.\,F$, where
$F$ is the Faraday tensor and ${}^{\ast}\!F$ its Hodge dual. Hence,
in this paper, we consider anyons theory in 3+1 dimensions and then the 
restriction onto the conducting layer.\par
Generally, this term is added up in order to consider monopoles such as
the Dirac one. Let us recall that the Dirac monopole is associated
with a  singular gauge potential. Hence people have made models of
anyons considering interaction between free electrons and magnetic monopoles.
As S. Mandelstam   showed \cite{Mandelstam}, this leads to a
confinement of the Fermi gas and may generate superconductor states. Anyway,
statistical calculus based on the equivalence between anyons gas and free
electrons embeded in a static magnetic field, show that the anyons gas
generates a Meissner effect. That is the main reason, with the
existence of permanent currents in the quantized Hall effect, for
using anyons theory in order to explain high-T$_c$ superconductivity
in two dimensions.

\section{\normalsize\hskip-.5cm - Toroidal moments and broken
relativistic crystalline symmetries}

Electric and magnetic crystals are characterized by their
magnetic groups which are subgroups of the so-called Shubnikov group
$O(3)1'$ (the time inversion is indicated by the ``\,\,$'$\,\," symbol).
Among the 122 magnetic groups, only 106 are compatible with the existence of
a linear or quadratic magnetoelectric effect.\par 

In the present paper, we consider relativistic symmetry group theory
in crystals \cite{11}. Therefore, we need, first, an extension from the 
Shubnikov group $O(3)1'$ to the group $O(1,3)$ in the Minkowski
space, and second and  more particularly,   transformations of
$O(1,3)$ leaving invariant polarization and magnetization vectors,
and generating a subgroup of the normalizer $N(G)$ of $G$ in
$O(1,3)$. This subgroup $G'$ may not be identified with the magnetic
group if $G$ leaves invariant a particular  non-vanishing velocity
vector. If such a vector exists and $G'\neq G$, one strictly speaks
about the relativistic crystalline symmetry $G'$. Only 31 magnetic
groups are compatible with the existence of a relativistic
crystalline symmetry.\par

The invariant non-vanishing velocity vectors can be linked with toroidal
moments from the point of view of magnetic symmetries, as it has been
already shown in previous papers (see \cite{12} for instance). The toroidal
moments
$\bf T$ are polar tensors which change sign under time inversion, like
velocity vectors  or current vectors of electric charges.  There are
refered to  the order parameter in toroidal phase transitions,
and involved, in particularly, in the superdiamagnetism of superconductors or
dielectric diamagnetic bodies containing densely packed atoms
(agregates)  \cite{12}.\par

In such systems, in the presence of spontaneous currents, there may exist
states for which the configuration of the associated currents has a
tore-shaped  solenoid with a winding and so a toroidal moment. For a
system with a dipolar toroidal moment density $\vec T$, the current
density equals 
\begin{equation}
{\vec j} = curl(curl({\vec T}))\,.
\label{curl}
\end{equation}
We can remark
on Figure 1 below
\begin{center}
\fbox{\includegraphics[bb=59 164 503 816,scale=.44]{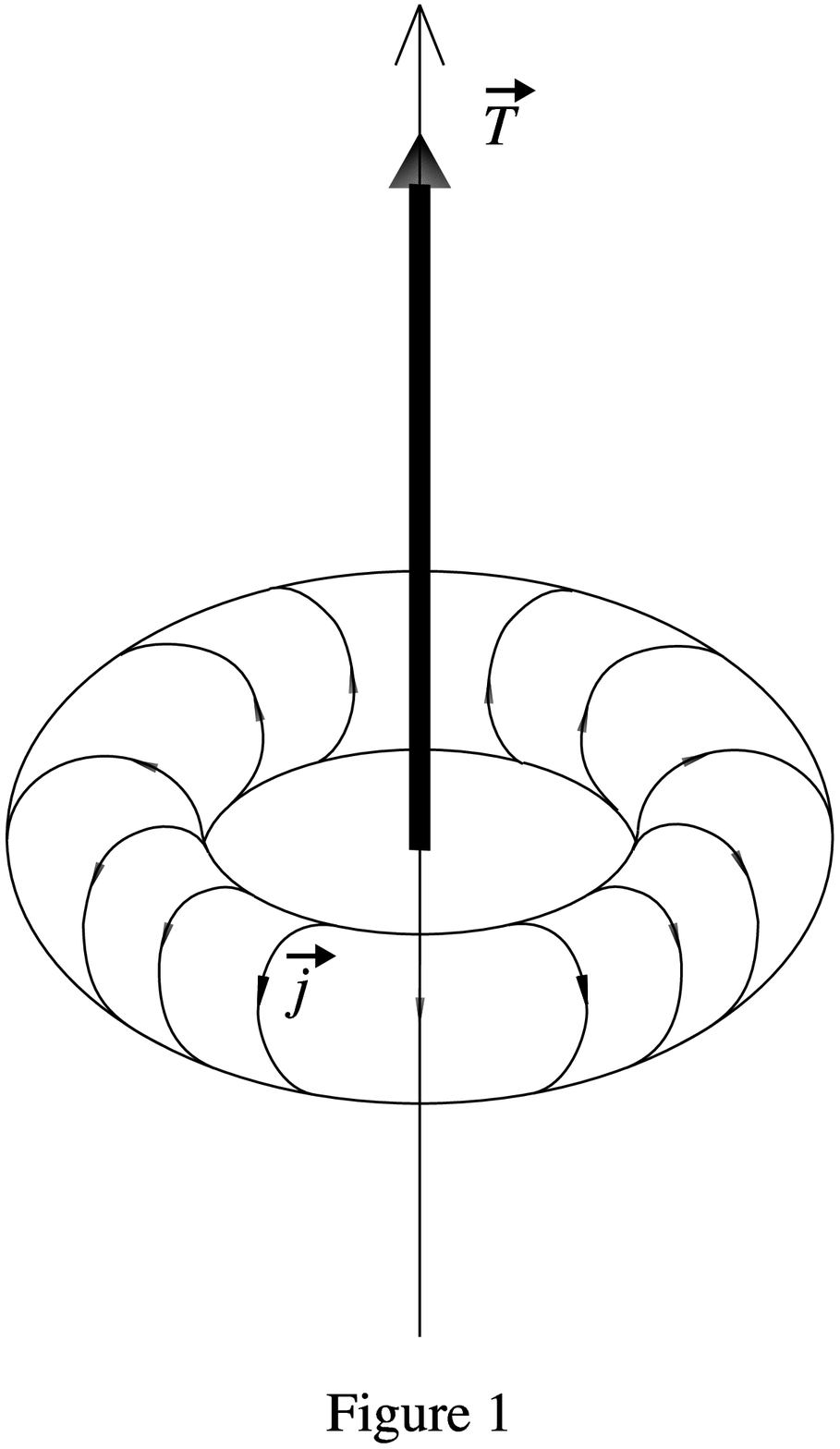}}
\end{center}
that this kind of configuration of currents can be
generated from a solenoidal one. The creation  or the closure to a
tore-shaped solenoid from the latter, can be obtained adiabatically, 
applying for instance, a homogeneous external magnetic field slowly rotating
at a more lower frequency than the cyclotron one. But we present another
possibility in connection with the breaking of the relativistic crystalline
symmetries. Before, it is absolutly necessary to understand that the
groups $G'$ are strongly depending on the orientations of the polarization
and/or magnetization vectors.\par  

Thus, if for example the system passes from the polarization vector ${\vec
P} = {\vec 0}$ to ${\vec P} \neq{\vec 0}$, it involves a 
breaking, by group conjugaison, of the relativistic crystalline symmetry.
Moreover, this conjugaison can be associated to a change of Galilean frame,
from an initialy fixed one to an ``effective" moving one, in such a way
that the initial $G'$ group is kept in the latter.\par

Then, phenomenologically, we can consider for instance, the following 
classical physical system:   a relativistic crystalline
group $G'_0$ of an elementary polarized cell of a given crystal, with an
electron not in interaction with this magnetic cell in the initial state. We
also assume that this electron has initially a Galilean motion with, for
instance, a velocity $\vec v$ parallel to the invariant axis of $G'_0$
(it is necessary to assume that this group is not invariant by smooth
translations in order to have only one such axis). Then, the
electron  interacts with the cell, inducing (by a local
magnetoelectric influence effect, by an exchange interaction for
instance or in fact whatever is the interaction !) a change of the  
electric or magnetic polarization which breaks the initial relativistic
crystalline symmetry group. During the interaction, the resulting symmetry
group is $G'_i$ in the fixed laboratory frame. The transformation
from one symmetry to the other is realized, by group conjugaison, with an
element $S$ of $N(G)$. It means that we pass with $S$ from one frame to
another in which the motion of the electron is kept, i.e. Galilean. In some
way, we can speak of a kind of ``inverse kineto-magnetoelectric effect",
since usually the crystal  moves whereas it is motionless in that
present case.\par

Thus, the motion of the electron in the laboratory frame is determined
by $S$ , i.e. at most a rotation around the invariant axis and a
boost along the same one. From the latter characteristics of the
electron motion and those of $S$, the electron will get a
solenoidal motion around the invariant axis, and if the electron is
polarized, the Thomas precession of its spin along the trajectory
would lead to a toroidal configuration of spin currents \cite{13}.
Moreover, if the electron has a fast solenoidal motion before the
interaction with the cell, then with the condition that the
interaction is adiabatic, one will obtain, by the latter  process,
a slow solenoidal motion of the mean position of the electron, and
so a tore-shaped motion during the interaction. Let us recall also that
a non trivial Berry phase can occur in this process with two main
consequences: an effective Yang-Mills field associated to an anomaly
such as a monopole and an anyonic statistic in cases of collective
motions \cite{14}.\par

Then, to finish with this description, we present the list of magnetic
groups compatible with such process of creation of toroidal
moments \cite{rubnuovo} (let us remark that among the 16 compatible groups
tabulated in \cite{rubnuovo}, only 12 are associated to a non-trivial
$O(2)$ action of the normalizer; that is why four of them, namely the
groups 2, 3, 4,  6, are not indicated in the list below): 
\[\begin{array}{l}
1\,,\,\, 2'\,,\,\, m\,,\,\, m'\,,\,\, {\bar 1}'\,,\,\, 2'\!/m\,,\,\,  {\bar
3}'\,,\\
\\
2\!/m'\,,\,\, 4\!/m'\,,\,\, 6\!/m'\,,\,\, {\bar 4}'\,,\,\, {\bar 6}'\,, 
\end{array}\]
(for $m$ and $2'$ we have two TM's oriented in an opposite direction; each
one corresponding to possible sublattices). All the
corresponding normalizers $N(G)$ contain, as a subgroup, the group of
rotations in the plane: $O(2)$, in order to have the precession phenomena.
To finish this chapter, let us give a supplementary important remark. The
electromagnetic and the ``anyonic" potential 4-vectors $\bf A$ are toroidal
moment densities, since  they satisfy both the relation (\ref{curl}) in a
3+1 dimensional anyons theory restricted in a 2+1 space, i.e.  the fields do
not depend on the perpendicular coordinate. Then in fact, the problem is  
to work out a non-electromagnetic potential vector in magnetoelectric
materials associated to each charge carrier. We will see, as a consequence,
that CS terms appear and so anyonic statistics. 

\section{\normalsize\hskip-.5cm - The relativistic effective Lagrangian}

In this chapter, we derive a density of free enthalpy, or
rather a relativistic Lagrangian density, since first, we are not
interest in thermal properties, and  second, because our aim is to
obtain a usable ``microscopic" Lagrangian in anyons theory.
Essentially for simplicity of notations and to avoid discussions
about meaning of a lot of particular tensorial coefficients, we will
consider a system in a ``nearly empty" space since the reasoning we
make, will be easily extended in matter without fundamental
modifications. The system we consider, is made of a free carrier and
only one polarized elementary magnetic cell freely moving in 
space. In fact, we consider a kind of local crystal field theory.
\par
In some way, the carrier with the cell is a polarized carrier,
meaning that the particle gets a covariant (invariant under application
of a boost and/or a rotation) tensorial property such as the following scalar
product ($\alpha = 0,1,2,3$ and Einstein convention):
\begin{equation}
{\bf v\,.\,u}\, =\,{v^\alpha}\,{u_\alpha}\, =\, cste\,,
\label{scalar}
\end{equation}
where $\bf{u}$ is the velocity
4-vector of the carrier (or of the mean position of the carrier in case
of an adiabatic process) and $\bf{v}$ is a
particular 4-vector associated to the magnetic cell. For instance, $\bf{v}$
can be the invariant velocity vector of the magnetic symmetry of the cell,
or its magnetic or electric polarization vectors. We assume that the 
interactions  between the polarization vectors  of the cell and the carrier 
are only depending on the relative carrier-cell position. As a fondamental
result of the necessary covariance  of the relativistic symmetry, any kind of
interaction  is equivalent to a change of Galilean frame and this is related
to the deep meaning of the concept of "polarized carriers". Here, we have a
carrier polarized by a cell (!) not only by a spin.\par
Then, the breaking occurs and the two 4-vectors $\bf u$ and $\bf v$ are
transformed by a Lorentz transformation ${\bf L}\subseteq N(G)$, i.e. we
observe in the laboratory frame the primed vectors ${\bf u' }={\bf L.u}$ and
${\bf v'} = {\bf L.v}$. Now, deriving relation (\ref{scalar}) with respect to
the laboratory frame time, we obtain:
\[{\bf\dot v\,.\,u} + {\bf v\,.\,{\dot u}}\, =\, 0\,,\]
where the dot indicates the time derivation. Decomposing ${\bf\dot u}$ in a
colinear and an orthogonal part to $\bf v$, we deduce (${\bf a\times b}\ =\
({a_\alpha}{b_\beta}-{a_\beta}{b_\alpha}){E_{\alpha,\beta}}$ where
$E_{\alpha,\beta}$ is the matrix with 1 on row $\alpha$ and column $\beta$
and zero everywhere else):
\[{\bf\dot u}\ =\ -({\bf{\dot v}\,.\,u})\,{\bf v} +  {\bf\dot u}_1=
\ ({\bf v\times{\dot v}}){\bf\,.\,u} + {\bf\dot u}_1\,.\]
It can be shown that ${\bf\dot u}_1$ is only due to different interactions
which don't break the symmetry \cite{15}. So, we cancel it out from the
latter expression.  \par
This little computation leads us to write:
\[{\bf\dot u}'\ =\ ({\bf\dot L}\,{\bf L}^{-1} - {\bf Lv\times L{\dot
v}})\,{\bf .\,u'}\
\equiv\ {\bf F}_{eff.}\,{\bf .\,u'}\,.\]
This equation is not but the least, a kind of generalized Thomas precession
equation \cite{15} which can be associated, in this particular system, to an
effective Faraday tensor ${\bf F}_{eff.}$ being only a function of the
relative cell-carrier position $\vec r$ and time, and having not 
an electromagnetic origin ! But, as a great surprise, the computation of
${\bf F}_{eff.}$ shows, first, that its  effective electric field is
vanishing, and second, its effective magnetic field is such that (${\bf v}
\equiv (\gamma,\gamma\, {\vec v}),
\gamma = (1-{\vec v}^2)^{-1/2}$; $\theta$ being a function of $t$ and $\vec
r$):
\begin{equation}
{\vec B}_{eff.} =
{m \over e}{\gamma^2 \over {(1+\gamma)}}\,{{\vec v}\, \wedge\, {\dot {\vec
v}}}\, \simeq\,\theta\,\, {{\vec v}\, \wedge\, ({\vec j}_e\,.\,{\vec
{\nabla}})\,{\vec v}}\,.
\label{beff}
\end{equation}
We think that it may be possible to interpret this
field as the effective magnetic field of the flux tubes leading to
the Aharanov--B\"ohm phase in the anyons theory. We can also
notice from the last term in (\ref{beff}) that ${\vec B}_{eff.}$
is a kind of vectorial Lifshitz invariant. From this expression, it
follows also that $\vec v$ can't be the invariant velocity vector
since the breaking doesn't affect its direction. So $\vec v$ is, up
to a suitable scalar or pseudoscalar factor, one of the two
polarization vectors of the magnetic cell. Thus, if for instance
$m'$ is the magnetic symmetry, since the two latter vectors are in
the plane of the symmetry, then ${\vec B}_{eff.}$ is in the
orthogonal direction. We have also to notice, as a consequence of 
the model, that there is such an effective magnetic field (and so a toroidal
moment) associated to each charge carrier.\par
At this step, unfortunately, we need very
sophisticated mathematical tools such as exterior differentiation $d$,
co-differentiation $\delta$, exact, co-exact and harmonic differential
n-forms, coming from the de Rham cohomology theory of differential
manifolds (see \cite{eguchi} for instance). \par
The fundamental consequence of the previous relation is that the
total Faraday tensor ${\bf F}_{Tot.} ={\bf F} + {\bf F}_{eff.}$ is
no longer a closed form, i.e. $d{\bf F}_{Tot.}\not =0$
($\Longleftrightarrow div{\vec B}_{eff.} \not = 0$) because of the time $t$
and the vector position 
$\vec r$ dependencies. It follows from the Hodge duality principle that  we
can define $d{\bf F}_{Tot.}$ ($\delta{\bf F}_{Tot.}$) as a  current of an
``effective" magnetic (electric) charge: $d{\bf F}_{Tot.} = {}^*\!{\bf j}_m$ 
($\delta{\bf F}_{Tot.} = {\bf j}_e$). Let us remark that the  currents are
not primed, because they are currents in the laboratory frame, or
equivalently, they have zero divergences in contradistinction with the 
primed currents. Moreover, it means also that the electric charge-matter
interaction is equivalent to an electric charge-magnetic charge interaction.
\par Hence, if the Lagrangian density for the ``electromagnetic" Faraday
tensor ${\bf F}$ is (with Heaviside's units):
\[-{1 \over 4} {\bf F\, .\, F}\, -\, {\bf A\, .\, j}\,,\]
we obtain for ${\bf F}_{Tot.}$ an other Lagrangian density  $\cal L$ such
that:
\begin{eqnarray}
&&\hskip-.4cm{\cal L}\equiv-{1 \over
4}\, {\bf F}_{Tot.}\, {\bf .}\, {\bf F}_{Tot.}\, -\, {\bf A}_e\, {\bf .}\,
{\bf j}_e\nonumber\\  &&\qquad\,\,\,- {\kappa \over 4}\,\,
^*{\bf F}_{Tot.}\, {\bf .}\, {\bf F}_{Tot.}\, -\, \kappa\, {\bf A}_m\,
{\bf .}\, {\bf j}_m\,,\qquad
\label{lag}
\end{eqnarray}
where $\kappa$ is a constant,  $^*{\bf F}_{Tot.}$ the Hodge dual of
${\bf F}_{Tot.}$, and ${\bf A}_e$ and ${\bf A}_m$ are respectively the
potential 4-vectors associated to the exact part of ${\bf F}_{Tot.}$
and $^*{\bf F}_{Tot.}$. As for the currents,  the ${\bf A}$'s are not
primed. Let us remark that the harmonic parts  have no contributions
to the dynamic of the charges, since they differentiations and
co-differentiations are vanishing, but they have one in the energy and
consequently in the Lagrangian density. These parts are associated to
the so-called instantons which are fields generated by the breaking
and  going away from the breaking area, carrying out a part of the
energy. Nevertheless, in the case of more than one electric charge,
this field will interact with the others and can not be neglected. \par
Thus, the total potential 4-vector $\bf A$, interacting with a one-charge
current, will be the sum of the potentials of all the closed parts of the
total corresponding Faraday tensor, for which the Poincar\'e's lemma can be
applied (indeed, from this lemma, in a suitable open subset not containing
the interaction area, it always exists a potential since there is no more
closed part inside). In fact, because  ${\bf F}_{Tot.}$ is not
only electromagnetic and
\[\begin{array}{l}
\delta{\bf F}_{Tot.} = {\bf j}_e
\Longrightarrow  curl(curl({\vec A}_e))\ \buildrel {def.}
\over \equiv\ {\vec j}_{{\bf A}_e}\\ 
d{\bf F}_{Tot.} = {}^*{\bf j}_m
\Longrightarrow   curl(curl({\vec A}_m))\ \buildrel {def.}
\over \equiv\ {\vec j}_{{\bf A}_m}\,,
\label{kapa}
\end{array}\] 
then ${\vec A}_e$ and ${\vec A}_m$ 
are toroidal moment densities: ${\vec A}_{e,m}\ \equiv\ {\vec
T}_{e,m}$. Moreover, $^*{\bf F}_{Tot.}\, {\bf .}\, {\bf F}_{Tot.}$ is the
expected Chern-Simon term of anyons theory and the harmonic part of
${\bf F}_{Tot.}$ will correspond to the famous singular potential
of this theory, i.e. to magnetic monopoles. Consequently, at a
macroscopic level, we must add in the free enthalpy, terms such as
($E$: electric field, $H$: magnetic field, $J$: current, $T$: dipolar
toroidal moment):
\[\begin{array}{l}
\gamma_{i,j}\, E^i\, J_{e,m}^j\,,\quad \mu_{i,j}\, H^i\, J_{e,m}^j\,,\\
\\
\rho_{i,j}\, J_{e,m}^i\, J_{e,m}^j\,,\quad \lambda_{i,j}\, J_e^i\, J_m^j\,,
\end{array}\]
terms, up to order four, coming from relations (\ref{beff}) and
(\ref{lag}), analogous terms substituting $T^i$ for $J^i$, and crossing
terms such as
\[\tau_{i,j} J_{e,m}^i T_{e,m}^j\,,\quad \omega_{i,j}  J_{e,m}^i
T_{m,e}^j\,,\] 
where $\gamma$, $\mu$, $\rho$, $\lambda$, $\tau$ and
$\omega$ are the corresponding suitable susceptibility tensors. To
finish, let us add that it might allow permanent currents
for particular choices of the latter tensors, and so superconductivity
with toroidal phase transitions~\cite{12}.

\end{document}